%% file: Demo__RTKiller_controlling_GNSS_rovers_by_RTK_base_station_spoofing.tex
\documentclass[sigconf,screen,nonacm]{acmart}

\acmConference[WiSec '24]{WiSec '24: 17th ACM Conference on Security and Privacy in Wireless and Mobile Networks}{May 27--May 30, 2024}{Seoul, Korea}
\acmBooktitle{WiSec '24: 17th ACM Conference on Security and Privacy in Wireless and Mobile Networks, May 27--May 30, 2024, Seoul, Korea}
\acmPrice{15.00}
\acmISBN{978-1-4503-XXXX-X/21/06}

\usepackage{glossaries}
\usepackage{siunitx}
\usepackage{caption}
\usepackage{subcaption}
\usepackage[capitalise]{cleveref}
\glsdisablehyper %% http://tug.ctan.org/macros/latex/contrib/glossaries/glossariesbegin.pdf -> disable highlighted gls entries in screen version, clickable ones for all versions
	
\input{acronyms.tex}

\begin{document}

\title{DEMO: RTKiller - manipulation of GNSS RTK rovers by reference base spoofing}
% if too long use  \title[short title]{full title}
%\subtitle{how to spoof gnss signals via replay}

\author{Marco Spanghero}
\orcid{0000-0001-8919-0098}
\affiliation{%
	\institution{Networked Systems Security Group \\ KTH Royal Institute of Technology}
	\streetaddress{Isafjordsgatan 26, 16440 Kista}
	\city{Stockholm}
	\country{Sweden}
}
\email{marcosp@kth.se}

\author{Panagiotis Papadimitratos}
\orcid{0000-0002-3267-5374}
\affiliation{%
	\institution{Networked Systems Security Group \\ KTH Royal Institute of Technology}
	\streetaddress{Isafjordsgatan 26, 16440 Kista}
	\city{Stockholm}
	\country{Sweden}
}
\email{papadim@kth.se}

%\renewcommand{\shortauthors}{Lenhart et al.}

\begin{abstract}
Global Navigation Satellite Systems (GNSS) provide global positioning and timing. Multiple receivers with known reference positions (stations) can assist mobile receivers (rovers) in obtaining GNSS corrections and achieve centimeter-level accuracy on consumer devices. However, GNSS spoofing and jamming, nowadays achievable with off-the-shelf devices, are serious threats to the integrity and robustness of public correction networks. In this demo, we show how manipulation of the \gls{pnt} solution at the reference station is reflected in the loss of baseline fix or degraded accuracy at the rover. \gls{rtk} corrections are valuable but fundamentally vulnerable: attacking the reference stations can harm all receivers (rovers) that rely on the targeted reference station.
\end{abstract}

\maketitle

\section{Introduction}
\gls{gnss} services are commonly used to provide precise timing and localization, globally, to a wide set of applications. Standalone GNSS can achieve sub-meter accuracy, but \gls{rtk} is necessary in high-accuracy contexts. \gls{rtk} allows using multiple GNSS receivers to obtain centimeter-level accuracy correcting measurement errors by comparing a moving receiver (rover) with a static one (station). \gls{rtk} is commonly used for \gls{uav} navigation, autonomous driving, robotics, and precision agriculture applications, as it is simple and cost-effective. Modern \gls{rtk} infrastructure relies on network-based \gls{rtk}, unlike traditional \gls{rtk} with short-range radios connecting the rover and the station receiver. This simplifies both the extension of and the accessibility to correction services. Several reference stations are connected to the Internet and any device can leverage network connectivity to obtain out-of-band \gls{gnss} corrections; inversely, a single reference station can provide corrections to multiple receivers, usually within \SI{10}{\kilo\meter} from the station location. 

\gls{rtk} reference stations are mounted at precisely surveyed points, whose location is accurately determined. This way, the reference receiver can calculate precise carrier phase measurements referenced to its static position and distribute this information to the connected rovers, which ultimately use them to calculate precise positions. The corrections are meaningful only if the rover can securely access the reference station (e.g., via an authenticated and encrypted link as in \cite{Pepjin:2020}), and the reference station itself is not under adversarial control. 
Nevertheless, each RTK reference receiver, even if otherwise trusted, can be jammed or spoofed: cryptographical protocols safeguarding the network-based correction distribution system cannot contain this problem. 

Spoofing any mobile GNSS receiver, in this context the rover, is complex, but it is much simpler and straightforward to smoothly capture a GNSS receiver whose location is well-determined and fixed. By attacking the RTK reference, the attacker is capable of causing significant degradation of the \gls{pnt} solution at the rover, resulting in potentially infeasible trajectory control or outright denial of service. This is specifically true in cases where RTK corrections are necessary to obtain the required accuracy and any rover that obtains corrections from a reference station under adversarial control data will be affected.

This is exactly the aim of this demo: under different \gls{gnss} constellation configurations and attack methods, we show that a strategically placed attacker can degrade the \gls{pnt} solution quality at the rover simply by tampering with the reference station receiver. We demonstrate three adversarial settings (synchronous single constellation lift-off, multiconstellation asynchronous spoofing, and jamming), and the effects they have on the rover's \gls{pnt} solution.

\section{System and Adversary model}

\begin{figure}[h!]
    \centering
    \includegraphics[width=\linewidth]{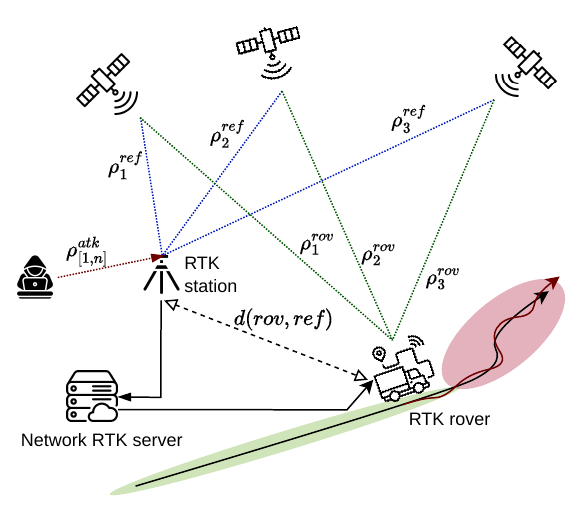}
    \caption{Typical network RTK scenario with an attacker spoofing the reference station}
    \label{fig:atk-pictorial}
\end{figure}

Two \gls{gnss} receivers, one configured as \gls{rtk} rover and the other as \gls{rtk} station, are connected over a network link. The \gls{rtk} station is mounted in a fixed point with known coordinates, which are included in the correction stream that is provided over the network. The network link between the rover and the station is protected by standard network practices (e.g., authenticated link). Practically, the adversary can only control the reference \gls{gnss} receiver by simulation, co-simulation, or replay-relay of \gls{gnss} signals. The attack setup is simplified in \cref{fig:atk-pictorial}, where the attacker causes a reduction of accuracy at the rover by the transmission of interference to the reference station.

Due to the open structure of the \gls{gnss} signal the adversary can craft signals with valid modulation, frequency, and data content that match the position of the \gls{rtk} station and, jointly, the adversarial action. This usually consists of (but is not limited to) modifications in the navigation data information, code-carrier modification, and selective replacement of specific signals \cite{humphreysAssessingSpoofingThreat2008, LenhartSP:C:2022}. Specifically, the adversary can match and synchronize the spoofing signals to the legitimate ones and slowly force its adversarial action. It is important to notice here that the attacker aims to disrupt and degrade the \gls{rtk} solution quality at the rover. In a regular spoofing scenario, the adversary usually cannot measure the effect of its action on the victim receiver. Due to the broadcast nature of the network-based \gls{rtk}, the attacker can monitor the degradation caused to the correction stream by tampering with the \gls{gnss} signals simply by connecting its own rover receiver to the connection stream.

\section{Experiment Setup}
\label{sec:setup} 
The \gls{rtk} rover implementation relies on RTKLIB for the baseline calculation and \gls{rtk} corrections, providing the user with a 3D corrected \gls{pnt} solution. The receivers on each \gls{gnss}-enabled node are u-Blox ZED-F9P high precision \gls{rtk} receivers. 

\begin{figure}[h!]
    \centering
    \includegraphics[width=0.7\linewidth]{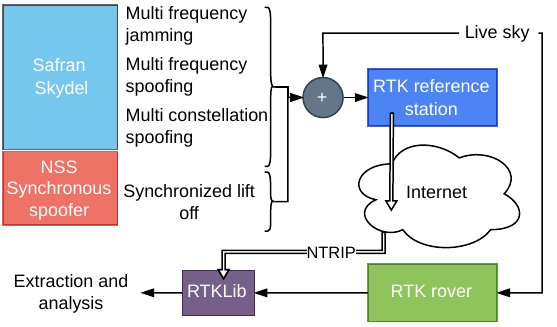}
    \caption{Demo setup.}
    \label{fig:experiment-pictorial}
\end{figure}

The adversary is implemented in two different ways. First, a custom-made synchronous signal simulator targeting L1 is used. This implementation allows the adversary to generate code phase-matched GPS L1 signals for each satellite consistent with the \gls{rtk} station's live sky view and transmit them synchronously to the beginning of the GPS frame at the victim receiver. The attacker can modify the pseudoranges at the victim receiver while maintaining the victim's fixed position consistent with the operational conditions of the \gls{rtk} station. Second, an asynchronous spoofing attack is mounted against the \gls{rtk} station, targeting multiple constellations and forcing the \gls{rtk} station into a different position. The second adversarial setting allows the transmission of different jamming signals and conditions detrimental to the station \gls{pnt} solution quality. The experimental setup of the demo is shown in \cref{fig:experiment-pictorial}. This demonstration will showcase the implementation of the attack using both our in-house simulator and Safran Skydel to generate a legitimate constellation and a spoofing signal for a station and a rover receiver. The attendees will have the possibility of directly interacting with the setup and testing our live-sky spoofer. 

%%=============================================
\section{Evaluation and conclusions}
\label{sec:eval}

The experiment shows that the 3D error under spoofing is \SI{30}{\meter}, with peaks of several 100s of meters and the baseline calculation fails in 47.12\% of the cases (where the GNSS receiver and RTKLib default to a differential solution, instead of a fixed RTK one). The investigation is ongoing to explore whether the manipulation of the reference could cause predictable adversary-induced behavior at the rover. Results of this work are presented also in \cite{SpangheroMPP:Poster:wisec24}

\begin{figure}[h!]
    \centering
    \includegraphics[width=0.9\linewidth]{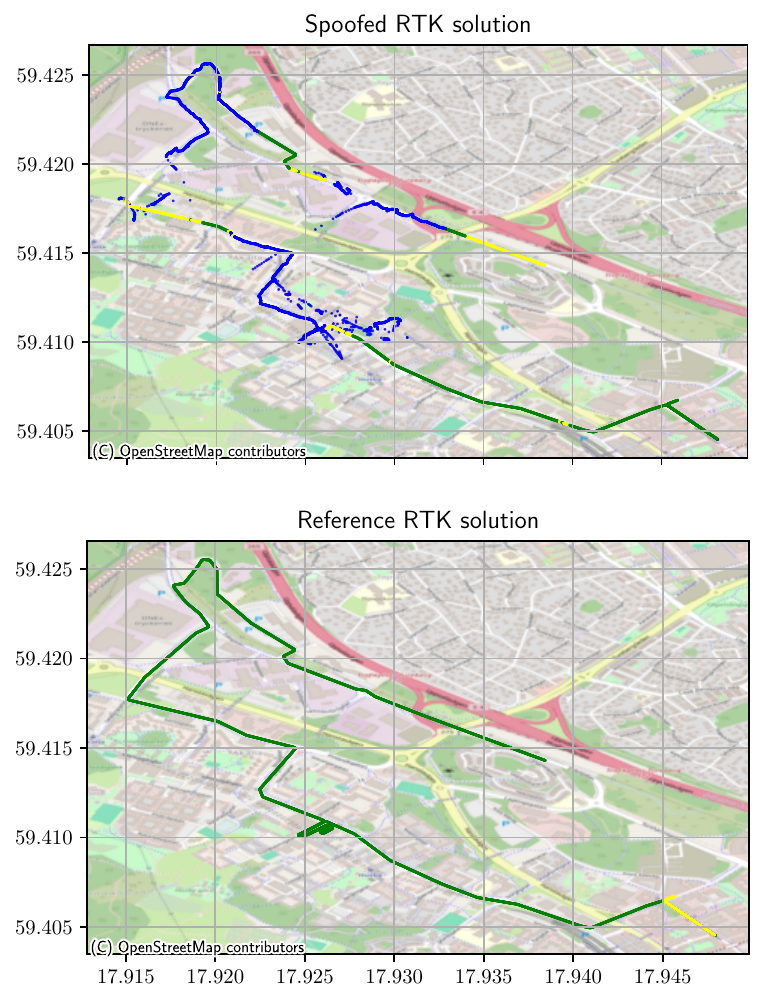}
    \caption{RTKiller in action causing degraded DGNSS solution at the rover.}
    \label{fig:rtkiller-result}
\end{figure}

\begin{acks}
    This work was supported by the Swedish Foundation for Strategic Research (SSF) SURPRISE project, the KAW Academy Fellow Trustworthy IoT project, and the Safran Minerva program. 
\end{acks}

\bibliographystyle{ACM-Reference-Format}
\bibliography{gnss-update}

\end{document}

%% file: acronyms.tex
\newacronym{gnss}{GNSS}{Global Navigation Satellite System}
\newacronym{gps}{GPS}{Global Positioning System}
\newacronym{nmea}{NMEA}{National Marine Electronics Association}
\newacronym{nasa}{NASA}{National Aeronautics and Space Administration}
\newacronym{agps}{A-GPS}{Assisted/Augmented GPS}
\newacronym{scer}{SCER}{Security Code Estimation and Replay}
\newacronym{nma}{OS-NMA}{Open Service Navigation Message Authentication}
\newacronym{sis}{SIS}{Signal In Space}
\newacronym{raim}{RAIM}{Receiver autonomous integrity monitoring}
\newacronym{pnt}{PNT}{Position-Navigation-Time}
\newacronym{rtk}{RTK}{Real Time Kinematics}

\newacronym{sdr}{SDR}{Software Defined Radio}
\newacronym{sdrs}{SDR}{Software Defined Radios}
\newacronym{ots}{OTS}{Off-the-Shelf}
\newacronym{rt}{RT}{Real-Time}
\newacronym{ide}{IDE}{Integrated Development Environment}
\newacronym{fifo}{FIFO}{First-In-First-Out}
\newacronym{sma}{SMA}{SubMiniature Version A}
\newacronym{rf}{RF}{Radio Frequency}
\newacronym{uav}{UAV}{Unmanned Aerial Vehicle}